\documentclass[prb,preprintnumbers,amsmath,amssymb,floatfix]{revtex4}

\usepackage{graphicx}
\usepackage{subfigure}
\usepackage{dcolumn}

\begin{document}
\title{Entanglement dynamics of electrons and photons}
\author{Xiang-Yao Wu$^{a}$ \footnote{E-mail: wuxy2066@163.com}, Xiao-Jing
Liu$^{a,b}$, Jing-Bin Lu$^{b}$ \\ Tian-Shun Li$^{a}$, Si-Qi
Zhang$^{b}$, Yu Liang$^{a}$, Ji Ma$^{a}$ and Hong Li$^{c}$}
 \affiliation{a. Institute of Physics, Jilin Normal
University, Siping 136000 China\\ b. Institute of Physics, Jilin
University, Changchun 130012 China\\ c. Institute of Physics,
Northeast Normal University, Changchun 130012 China}

\begin{abstract}
Entanglement is a fundamental feature of quantum theory as well as
a key resource for quantum computing and quantum communication,
but the entanglement mechanism has not been found at present. We
think when the two subsystems exist interaction directly or
indirectly, they can be in entanglement state. such as, in the
Jaynes-Cummings model, the entanglement between an atom and light
field comes from their interaction. In this paper, we have studied
the entanglement mechanism of electron-electron and photon-photon,
which are from the spin-spin interaction. We found their total
entanglement states are relevant both space state and spin state.
When two electrons or two photons are far away, their entanglement
states should be disappeared even if their spin state is
entangled.
\end{abstract}
\maketitle

{\bf 1. Introduction} \vskip 8pt

One of the most peculiar properties of quantum mechanics is
entanglement, that is the possibility to construct quantum states
of several subsystems that cannot be factorized into a product of
individual states of each subsystem. Such entangled states are the
most common in quantum mechanics, and they display correlations
which cannot be seen in a classical world.

Quantum entanglement has been extensively studied in the past few
years, both due to its fundamental significance in quantum theory
[1, 2], and it is the basic concept of the quantum information
processes, such as quantum computing [3], quantum teleportation
[4], quantum cryptography [5] and quantum communication [6]. The
atom-photon entanglement has been studied in atomic cascade
systems [7, 8] as well as in trapped ions [9, 10]. The observation
of the quantum entanglement between a single trapped $87Rb$ atom
and a single photon at a wavelength suitable for low-loss
communication has been reported [11]. Theoretical description of
entanglement evolution between atom and quantized field in the
Jaynes-Cummings model has been proposed [12-14]. However, it was
shown that the induced entanglement between two interacting
two-level quantum systems can be controlled by the relative phase
of applied fields [15]. In another study it was shown that the
atom-photon entanglement near a 3D anisotropic photonic band edge
depends on the relative phase of applied fields [16].

Entangled states of photons are the basic resource in the
successful implementation of quantum information processing
applications. The standard method for generating entangled photon
states nowadays is spontaneous parametric down conversion (SPDC),
which is achieved by pumping one or more nonlinear crystals with a
laser source. could have important benefits to applications in
optical quantum information. Photonic quantum gates require pure
states, which can be created by heralded sources producing pairs
of spectrally decorrelated photons [17, 18]. On the other hand,
long distance fiber based quantum communication and quantum
metrology suffers from chromatic dispersion, which could
potentially be improved with positive spectral correlations [19,
20].

In this paper, we have studied the entanglement mechanism of
electron-electron and photon-photon. With the spin-spin
interaction, we have given the spin entanglement states of two
electrons, three electrons, two photons, and found their total
entanglement states are relevant both space state and spin state.
When two electrons or two photons are far away, their entanglement
states should be disappeared even if their spin state is
entangled.

 \vskip 5pt {\bf 2. The entanglement between an atom and light field} \vskip 5pt

Entanglement can not be produce in isolation, when the two
subsystems exist interaction directly or indirectly, they can be
in entanglement state. In Jaynes-Cummings model, the entanglement
between an atom and light field is from their interaction. So what
interaction lead to the entanglement of electron-electron and
photon-photon? we think their spin-spin interaction form their
entanglement. In the following, we should prove that the
electron-electron and photon-photon spin entanglement states come
from their spin-spin interaction. \\

For the two-photon Jaynes-Cummings model, the Hamiltonian is [21,
22]

\begin{eqnarray}
H=\omega
a^{+}a+\frac{1}{2}\omega_{0}\sigma_{z}+g({a^{+}}^{2}\sigma_{-}+a^{2}\sigma_{+}),\hspace{0.2in}(\hbar=1)
\end{eqnarray}
describing the interaction of a light field with a two-level atom,
where $\omega$, $\omega_0$ are the frequencies of light field and
atom, $a (a^+)$ denotes the annihilation (creation) operator of
light field, $\sigma_{z}=\mid a><a\mid-\mid b><b\mid$,
$\sigma_{+}=\mid a><b\mid$, $\sigma_{-}=\mid b><a\mid$ are the
atomic operators, $\mid b>$ ($\mid a>$) is the ground (excited)
state of the atom, and $g$ is the coupling strength of atom and
light field, which reflects the interaction intensity of atom and light field.\\
At the initial time, the atom is at excited state and the
quantized field photon number is $N$, the initial state is
\begin{eqnarray}
\mid\psi(0)>=\mid a,n>,
\end{eqnarray}
with Schrodinger equation, we can obtain the state at any time
$t$, it is
\begin{eqnarray}
\mid\psi(t)>=c_1(t)\mid a,n>+c_2(t)\mid b,n+2>,
\end{eqnarray}
where
\begin{eqnarray}
c_1(t)=e^{-i\omega(n+1)t}[\cos(\frac{\sqrt{\delta^2+\omega_1^2}}{2}t)+
i\delta(\delta^2+\omega_1^2)^{-1/2}\sin(\frac{\sqrt{\delta^2+\omega_1^2}}{2}t)],
\end{eqnarray}
\begin{eqnarray}
c_2(t)=-ie^{i\omega(n+1)t}\omega_1(\delta^2+\omega_1^2)^{-1/2}\sin(\frac{\sqrt{\delta^2+\omega_1^2}}{2}t),
\end{eqnarray}
where $\delta=\omega_0-2\omega$ is the mismatching quantity and
$\omega_1=2g\sqrt{(n+1)(n+2)}$ is the Rabi frequency.\\
From Eq. (5), we can find when the interaction intensity of atom
and light field $g=0$, the coefficient $c_2(t)=0$, the
entanglement of atom and light field should be disappeared. So,
the entanglement of atom and light field is from the the
interaction of atom and light field, i.e., the interaction between
particles generate the entanglement between the particles.
Similarly, the entanglement between the electrons as well as the
photons come from the interaction between the electrons and the
photons. In the following, we shall study the entanglement
dynamics between the electrons and the photons.\\

\vskip 5pt {\bf 2. The entanglement between the teo electrons}
\vskip 5pt

The two electrons Hamiltonian operator is
\begin{eqnarray}
\hat{H}&=&-\frac{-\hbar^2}{2m}\nabla_1^2-\frac{-\hbar^2}{2m}\nabla_1^2+V(r_1,
r_2)+g\vec{s_1}\cdot \vec{s_2} \nonumber\\
&=&\hat{H_1}+\hat{H_2}+\hat{H_I}(r_1, r_2)+\hat{H_I}(s_1, s_2),
\end{eqnarray}
where $\hat{H_1}=-\frac{-\hbar^2}{2m}\nabla_1^2$,
$\hat{H_2}=-\frac{-\hbar^2}{2m}\nabla_2^2$, $\hat{H_I}(r_1,
r_2)=V(r_1, r_2)$, $\hat{H_I}(s_1, s_2)=g_e\vec{s_1}\cdot
\vec{s_2}$, $V(r_1, r_2)$ is the potential
energy£¬$g_e\vec{s_1}\cdot \vec{s_2}$ is the spin-spin interaction
energy of two electrons, $g_e$ is electrons spin coupling
strength, $\vec{s_1}$ and
$\vec{s_2}$ are the spins of electron $1$ and $2$.\\
The eigen equation of Eq. (6) is
\begin{equation}
(\hat{H_1}+\hat{H_2}+\hat{H_I}(r_1, r_2)+\hat{H_I}(s_1,
s_2))\psi(r_1, r_2, s_{1z}, s_{2z})=E\psi(r_1, r_2, s_{1z},
s_{2z}),
\end{equation}
where $E$ is the total energy of two electrons, the Eq. (7) can be
solved by separation variable method, it is
\begin{equation}
\psi(r_1, r_2, s_{1z}, s_{2z})=\psi(r_1, r_2)\cdot\chi_{SM_s}
(s_{1z}, s_{2z}),
\end{equation}
substituting Eq. (8) into (7), we can obtain
\begin{equation}
(\hat{H_1}+\hat{H_2}+\hat{H_I}(r_1, r_2))\psi(r_1,
r_2)=E_r\psi(r_1, r_2),
\end{equation}
and
\begin{equation}
\hat{H_I}(s_1, s_2)\chi_{SM_s} (s_{1z}, s_{2z})=g_e\vec{s_1}\cdot
\vec{s_2}\chi_{SM_s} (s_{1z}, s_{2z})=E_s\chi_{SM_s} (s_{1z},
s_{2z}),
\end{equation}
where $E_s=E-E_r$, the space and spin wave functions of two
electrons can be solved by Eq. (9) and (10), respectively. So, the
spin entanglement states of two electrons can be obtained by Eq.
(10).\\

Since
$\vec{s_1}\cdot\vec{s_2}=\frac{1}{2}(\vec{S}^2-\frac{3\hbar^2}{2})$,
the Eq. (10) can be written as
\begin{equation}
\frac{1}{2}g_e(\vec{S}^2-\frac{3\hbar^2}{2})\chi_{SM_s} (s_{1z},
s_{2z})=E_s\chi_{SM_s} (s_{1z}, s_{2z}),
\end{equation}
where $\vec{S}=\vec{s_1}+\vec{s_2}$ is the total spin of the two
electrons. In quantum mechanics, the Eq. (11) has four spin
eigenfunctions, they are

(1) $S=1$ two electrons spin symmetrical states
\begin{eqnarray}
\chi_{SM_s}^{(1)}=\chi_{\frac{1}{2}}(s_{1z})\chi_{\frac{1}{2}}(s_{2z})\hspace{0.3in}S=1,\hspace{0.1in}M_{s}=1,
\end{eqnarray}
\begin{eqnarray}
\chi_{SM_s}^{(2)}=\chi_{-\frac{1}{2}}(s_{1z})\chi_{-\frac{1}{2}}(s_{2z})\hspace{0.3in}S=1,\hspace{0.1in}M_{s}=-1,
\end{eqnarray}
\begin{eqnarray}
\chi_{SM_s}^{(3)}=\frac{1}{\sqrt{2}}[\chi_{\frac{1}{2}}(s_{1z})\chi_{-\frac{1}{2}}(s_{2z})
+\chi_{\frac{1}{2}}(s_{2z})\chi_{-\frac{1}{2}}(s_{1z})]
\hspace{0.3in}S=1,\hspace{0.1in}M_{s}=0,
\end{eqnarray}
(2) $S=0$ two electrons spin antisymmetrical state
\begin{eqnarray}
\chi_{SM_s}^0=\frac{1}{\sqrt{2}}[\chi_{\frac{1}{2}}(s_{1z})\chi_{-\frac{1}{2}}(s_{2z})
-\chi_{\frac{1}{2}}(s_{2z})\chi_{-\frac{1}{2}}(s_{1z})]
\hspace{0.3in}S=0,\hspace{0.1in}M_{s}=0.
\end{eqnarray}
where $M_{s}=s_{1z}+s_{2z}$. The states $\chi_{SM_s}^{(3)}$ and
$\chi_{SM_s}^{(0)}$ are the spin entanglement states of two electrons.\\

\vskip 5pt {\bf 3. The entanglement between the three electrons}
\vskip 5pt

For the three electrons, their Hamiltonian operator of electrons
spin interaction is
\begin{eqnarray}
\hat{H_I}(s_1, s_2, s_3)=g_e(\vec{s_1}\cdot
\vec{s_2}+\vec{s_2}\cdot \vec{s_3}+\vec{s_3}\cdot \vec{s_1}).
\end{eqnarray}
In three electrons system, the spin of two-electron and
three-electron $\vec{S}_{12}$ and $\vec{S}$ are
\begin{equation}
\vec{S}_{12}=\vec{s}_{1}+\vec{s}_{2},
\end{equation}
and
\begin{equation}
\vec{S}=\vec{s}_{1}+\vec{s}_{2}+\vec{s}_{3}=\vec{S}_{12}+\vec{s}_{3},
\end{equation}
their square are
\begin{equation}
\vec{S}_{12}^{2}=\frac{3}{2}\hbar^2+2\vec{s}_{1}\cdot\vec{s}_{2},
\end{equation}
and
\begin{eqnarray}
\vec{S}^{2}=\frac{9}{4}\hbar^2+2(\vec{s}_{1}\cdot\vec{s}_{2}+\vec{s}_{2}\cdot\vec{s}_{3}+\vec{s}_{3}\cdot\vec{s}_{1}),
\end{eqnarray}
substituting Eq. (20) into (16), we have
\begin{eqnarray}
\hat{H_I}(s_1,s_2,s_3)=\frac{g_e}{2}(\vec{S}_{12}^{2}-\frac{9}{4}\hbar^2),
\end{eqnarray}
the operators $\{\vec{S}^{2},\vec{S'}^{2},s_z\}$ common
eigenfunctions $\chi_{S'SM_s}$ are the eigenfunctions of Eq. (21),
and the corresponding quantum numbers $S'$ and $S$ are taken as
\begin{eqnarray}
S=\frac{3}{2},\hspace{0.3in}S'=1,
\end{eqnarray}
\begin{eqnarray}
S=\frac{1}{2},\hspace{0.3in}S'=1,0.
\end{eqnarray}
In quantum mechanics, the Eq. (21) has eight spin eigenfunctions,
they are

(1) $S\hspace{0.02in}'=1$, $S=\frac{3}{2}$ spin wave functions
$\chi_{S'SM_s}$
\begin{eqnarray}
\chi_{1\frac{3}{2}\frac{3}{2}}=\chi_{\frac{1}{2}}(s_{1z})\chi_{\frac{1}{2}}(s_{2z})\chi_{\frac{1}{2}}(s_{3z}),
\end{eqnarray}
\begin{eqnarray}
\chi_{1 \frac{3}{2}
\frac{1}{2}}&=&\frac{1}{\sqrt{3}}[\chi_{-\frac{1}{2}}(s_{1z})\chi_{\frac{1}{2}}(s_{2z})\chi_{\frac{1}{2}}(s_{3z})
+\chi_{\frac{1}{2}}(s_{1z})\chi_{-\frac{1}{2}}(s_{2z})\chi_{\frac{1}{2}}(s_{3z})
\nonumber\\&&+\chi_{-\frac{1}{2}}(s_{1z})\chi_{\frac{1}{2}}(s_{2z})\chi_{-\frac{1}{2}}(s_{3z})],
\end{eqnarray}
\begin{eqnarray}
\chi_{1 \frac{3}{2}
-\frac{1}{2}}&=&\frac{1}{\sqrt{3}}[\chi_{-\frac{1}{2}}(s_{1z})\chi_{-\frac{1}{2}}(s_{2z})\chi_{\frac{1}{2}}(s_{3z})
+\chi_{-\frac{1}{2}}(s_{1z})\chi_{\frac{1}{2}}(s_{2z})\chi_{-\frac{1}{2}}(s_{3z})
\nonumber\\&&+\chi_{\frac{1}{2}}(s_{1z})\chi_{-\frac{1}{2}}(s_{2z})\chi_{-\frac{1}{2}}(s_{3z})],
\end{eqnarray}
\begin{eqnarray}
\chi_{1 \frac{3}{2}
-\frac{3}{2}}=\chi_{-\frac{1}{2}}(s_{1z})\chi_{-\frac{1}{2}}(s_{2z})\chi_{-\frac{1}{2}}(s_{3z}).
\end{eqnarray}
(2) $S\hspace{0.02in}'=1$, $S=\frac{1}{2}$ spin wave functions
$\chi_{S'SM_s}$
\begin{eqnarray}
\chi_{1\frac{1}{2}\frac{1}{2}}&=&\frac{1}{\sqrt{6}}[2\chi_{\frac{1}{2}}(s_{1z})\chi_{\frac{1}{2}}(s_{2z})\chi_{-\frac{1}{2}}(s_{3z})
-\chi_{\frac{1}{2}}(s_{1z})\chi_{-\frac{1}{2}}(s_{2z})\chi_{\frac{1}{2}}(s_{3z})\nonumber\\
&&-\chi_{-\frac{1}{2}}(s_{1z})\chi_{\frac{1}{2}}(s_{2z})\chi_{\frac{1}{2}}(s_{3z})],
\end{eqnarray}
\begin{eqnarray}
\chi_{1\frac{1}{2}-\frac{1}{2}}&=&\frac{1}{\sqrt{6}}[\chi_{\frac{1}{2}}(s_{1z})\chi_{-\frac{1}{2}}(s_{2z})\chi_{-\frac{1}{2}}(s_{3z})
+\chi_{-\frac{1}{2}}(s_{1z})\chi_{\frac{1}{2}}(s_{2z})\chi_{-\frac{1}{2}}(s_{3z})
\nonumber\\&&-2\chi_{-\frac{1}{2}}(s_{1z})\chi_{-\frac{1}{2}}(s_{2z})\chi_{\frac{1}{2}}(s_{3z})].
\end{eqnarray}
(3) $S\hspace{0.02in}'=0$, $S=\frac{1}{2}$ spin wave functions
$\chi_{S'SM_s}$
\begin{eqnarray}
\chi_{0\frac{1}{2}\frac{1}{2}}=\frac{1}{\sqrt{2}}[\chi_{\frac{1}{2}}(s_{1z})\chi_{-\frac{1}{2}}(s_{2z})\chi_{\frac{1}{2}}(s_{3z})
-\chi_{-\frac{1}{2}}(s_{1z})\chi_{\frac{1}{2}}(s_{2z})\chi_{\frac{1}{2}}(s_{3z})],
\end{eqnarray}
\begin{eqnarray}
\chi_{0\frac{1}{2}-\frac{1}{2}}=\frac{1}{\sqrt{2}}[\chi_{\frac{1}{2}}(s_{1z})\chi_{-\frac{1}{2}}(s_{2z})
\chi_{-\frac{1}{2}}(s_{3z})-\chi_{-\frac{1}{2}}(s_{1z})\chi_{\frac{1}{2}}(s_{2z})\chi_{-\frac{1}{2}}(s_{3z})].
\end{eqnarray}
Obviously, the three electrons spin states
$\chi_{1\frac{3}{2}\frac{1}{2}}$,
$\chi_{1\frac{3}{2}-\frac{1}{2}}$,
$\chi_{1\frac{1}{2}\frac{1}{2}}$ and
$\chi_{1\frac{1}{2}-\frac{1}{2}}$ are their spin entanglement
states.

\vskip 5pt {\bf 4. The entanglement between the two photons}
\vskip 5pt

In section 2, the entanglement between the electrons come from
their Hamiltonian operator of electrons spin interaction. For the
two photons spin entanglement, we can assume they are from the
Hamiltonian operator of photons spin interaction, they are
\begin{equation}
\hat{H_I}(s_1, s_2)=g_{\gamma}\vec{s_1}\cdot \vec{s_2},
\end{equation}
where $g_{\gamma}$ is photons spin coupling strength, $\vec{s_1}$
and $\vec{s_2}$ are the spins of photon $1$ and $2$. The single
photon spin quantum number $s=1$, and the total spin square of
two-photon is
\begin{eqnarray}
\vec{S}\hspace{0.05in}^{2}=(\vec{s}_{1}+\vec{s}_{2})^{2}&=&\vec{s}_{1}\hspace{0.005in}^{2}+
\vec{s}_{2}\hspace{0.005in}^{2}+2\vec{s}_{1}\cdot
\vec{s}_{2}\nonumber\\&=& 4+2\vec{s}_{1}\cdot \vec{s}_{2},
\end{eqnarray}
and the total spin quantum numbers $S$ are
\begin{equation}
S=0,1,2.
\end{equation}
the Eq. (32) can be written as
\begin{equation}
\hat{H_I}(s_1,s_2)=\frac{1}{2}g_{\gamma}(\vec{S}\hspace{0.05in}^{2}-4),
\end{equation}
the spin eigenequation of two photons is
\begin{equation}
\frac{1}{2}g_{\gamma}(\vec{S}\hspace{0.05in}^{2}-4)\chi_{SM_s}=E_s\chi_{SM_s}.
\end{equation}
In Ref. [21], we have given the spin eigenequation of two photons,
they are

(1) $S=2$ two-photon spin symmetrical states
\begin{equation}
\chi_{22}=\chi_{1}(s_{1z})\chi_{1}(s_{2z}),
\end{equation}
\begin{equation}
\chi_{21}=\frac{1}{\sqrt{2}}[\chi_{0}(s_{1z})\chi_{1}(s_{2z})+\chi_{0}(s_{2z})\chi_{1}(s_{1z})],
\end{equation}
\begin{equation}
\chi_{20}=\frac{1}{\sqrt{6}}[\chi_{1}(s_{1z})\chi_{-1}(s_{2z})+2\chi_{0}(s_{1z})\chi_{0}(s_{2z})+\chi_{-1}(s_{1z})\chi_{1}(s_{2z})],
\end{equation}
\begin{equation}
\chi_{2-1}=\frac{1}{\sqrt{2}}[\chi_{0}(s_{1z})\chi_{-1}(s_{2z})+\chi_{-1}(s_{1z})\chi_{0}(s_{2z})],
\end{equation}
\begin{equation}
\chi_{2-2}=\chi_{-1}(s_{1z})\chi_{-1}(s_{2z}).
\end{equation}
(2) $S=1$ two-photon spin antisymmetrical states
\begin{equation}
\chi_{11}=\frac{1}{\sqrt{2}}[\chi_{1}(s_{1z})\chi_{0}(s_{2z})-\chi_{0}(s_{1z})\chi_{1}(s_{2z})],
\end{equation}
\begin{equation}
\chi_{10}=\frac{1}{\sqrt{2}}[\chi_{1}(s_{1z})\chi_{-1}(s_{2z})-\chi_{-1}(s_{1z})\chi_{1}(s_{2z})],
\end{equation}
\begin{equation}
\chi_{1-1}=\frac{1}{\sqrt{2}}[\chi_{0}(s_{1z})\chi_{-1}(s_{2z})-\chi_{-1}(s_{1z})\chi_{0}(s_{2z})].
\end{equation}
(3) $S=0$ two-photon symmetrical spin states
\begin{equation}
\chi_{00}=\frac{1}{\sqrt{3}}[\chi_{1}(s_{1z})\chi_{-1}(s_{2z})-\chi_{0}(s_{1z})\chi_{0}(s_{2z})+\chi_{-1}(s_{1z})\chi_{1}(s_{2z})],
\end{equation}
the single photon spin states $\chi_{0}$, $\chi_{1}$ and
$\chi_{-1}$ are [23]

\begin{eqnarray}
\chi_{0}= \left ( \begin{array}{lll}
   0  \\
   0 \\
   1  \\
   \end{array}
   \right ), \hspace{0.2in}
\chi_{1}= -\frac{1}{\sqrt{2}}\left ( \begin{array}{lll}
   1   \\
   i \\
   0  \\
   \end{array}
   \right ), \hspace{0.2in}
\chi_{-1}=\frac{1}{\sqrt{2}}\left ( \begin{array}{lll}
   1    \\
   -i  \\
   0   \\
   \end{array}
   \right ).
\end{eqnarray}
Obviously, the three photons spin states $\chi_{21}$, $\chi_{20}$,
$\chi_{2-1}$, $\chi_{11}$, $\chi_{10}$, $\chi_{1-1}$ and
$\chi_{00}$ are their spin entanglement states.

\vskip 5pt {\bf 5. The entanglement between the three photons}
\vskip 5pt

For the three photons, the Hamiltonian operator of their spin
interaction are
\begin{equation}
\hat{H_I}(s_1, s_2)=g_{\gamma}(\vec{s_1}\cdot
\vec{s_2}+\vec{s_1}\cdot \vec{s_3}+\vec{s_2}\cdot \vec{s_3}),
\end{equation}
the spin of two-photon and three-photon $\vec{S}_{12}$ and
$\vec{S}_{123}$ are
\begin{equation}
\vec{S}_{12}=\vec{s}_{1}+\vec{s}_{2},
\end{equation}
and
\begin{equation}
\vec{S}_{123}=\vec{s}_{1}+\vec{s}_{2}+\vec{s}_{3}=\vec{S}_{12}+\vec{s}_{3},
\end{equation}
their square are
\begin{equation}
\vec{S}_{12}^{2}=(\vec{s}_{1}+\vec{s}_{2})^{2}=\vec{s}_{1}\hspace{0.002in}^{2}+\vec{s}_{2}\hspace{0.002in}^{2}+2\vec{s}_{1}\cdot\vec{s}_{2}=4+2\vec{s}_{1}\cdot\vec{s}_{2},
\end{equation}
and
\begin{eqnarray}
\vec{S}_{123}^{2}=(\vec{s}_{1}+\vec{s}_{2}+\vec{s}_{3})^{2}&=&\vec{s}_{1}\hspace{0.002in}^{2}+\vec{s}_{2}\hspace{0.0002in}^{2}+\vec{s}_{3}\hspace{0.0002in}^{2}
+2(\vec{s}_{1}\cdot\vec{s}_{2}+\vec{s}_{2}\cdot\vec{s}_{3}+\vec{s}_{3}\cdot\vec{s}_{1})\nonumber\\
&=&6+2(\vec{s}_{1}\cdot\vec{s}_{2}+\vec{s}_{2}\cdot\vec{s}_{3}+\vec{s}_{3}\cdot\vec{s}_{1}),
\end{eqnarray}
obviously, $\vec{S}_{12}^{2}$ commutes with $\vec{S}_{12}$ and
$\vec{s}_{3}$, and $\vec{S}_{12}^{2}$ commutes with
$\vec{S}_{123}$. So, $\vec{S}_{12}^{2}$, $\vec{S}_{123}^{2}$ and
$(\vec{S}_{123})_z$ commute with each other, and they have common
eigenfunctions. The eigenvalues and quantum number of
$\vec{S}_{12}^{2}$ and $\vec{S}_{123}^{2}$ are
\begin{equation}
\vec{S}_{12}^{2}=S'(S'+1),\hspace{0.3in} S'=2,1,0,
\end{equation}
and
\begin{equation}
\vec{S}_{123}^{2}=S(S+1),\hspace{0.1in} S=3,2,1
(S\hspace{0.002in}'=2); \hspace{0.1in}2,1,0
(S\hspace{0.002in}'=1); \hspace{0.1in}1 (S\hspace{0.002in}'=0).
\end{equation}
By calculation, we can obtain the three photons spin eigenstates
and spin entanglement states, they are
\begin{equation}
\chi_{233}=\chi_{22}(s_{1z},s_{2z})\chi_{1}(s_{3z})=\chi_{1}(s_{1z})\chi_{1}(s_{2z})\chi_{1}(s_{3z}),
\end{equation}
the spin state $\chi_{233}$ is not the three-photon spin
entanglement state.\\
\begin{eqnarray}
\chi_{232}=\frac{1}{\sqrt{3}}[\chi_{0}(s_{1z})\chi_{1}(s_{2z})\chi_{1}(s_{3z})+\chi_{1}(s_{1z})\chi_{0}(s_{2z})\chi_{1}(s_{3z})+\chi_{1}(s_{1z})\chi_{1}(s_{2z})\chi_{0}(s_{3z})],
\end{eqnarray}
the spin state $\chi_{232}$ is the three-photon spin
entanglement state.\\
\begin{eqnarray}
\chi_{231}&=&\frac{1}{\sqrt{15}}[\chi_{-1}(s_{1z})\chi_{1}(s_{2z})\chi_{1}(s_{3z})+2\chi_{0}(s_{1z})\chi_{0}(s_{2z})\chi_{1}(s_{3z})\nonumber\\
&&+2\chi_{0}(s_{1z})\chi_{1}(s_{2z})\chi_{0}(s_{3z})+2\chi_{1}(s_{1z})\chi_{0}(s_{2z})\chi_{0}(s_{3z})\nonumber\\
&&+\chi_{1}(s_{1z})\chi_{-1}(s_{2z})\chi_{1}(s_{3z})+\chi_{1}(s_{1z})\chi_{1}(s_{2z})\chi_{-1}(s_{3z})],
\end{eqnarray}
the spin state $\chi_{231}$ is the three-photon spin
entanglement state.\\
\begin{eqnarray}
\chi_{230}&=&\frac{1}{\sqrt{10}}[\chi_{-1}(s_{1z})\chi_{1}(s_{2z})\chi_{0}(s_{3z})+2\chi_{0}(s_{1z})\chi_{0}(s_{2z})\chi_{0}(s_{3z})\nonumber\\
&&+\chi_{0}(s_{1z})\chi_{-1}(s_{2z})\chi_{1}(s_{3z})+\chi_{0}(s_{1z})\chi_{1}(s_{2z})\chi_{-1}(s_{3z})\nonumber\\
&&+\chi_{-1}(s_{1z})\chi_{0}(s_{2z})\chi_{1}(s_{3z})+\chi_{1}(s_{1z})\chi_{-1}(s_{2z})\chi_{0}(s_{3z})\nonumber\\
&&+\chi_{1}(s_{1z})\chi_{0}(s_{2z})\chi_{-1}(s_{3z})],
\end{eqnarray}
the spin state $\chi_{230}$ is the three-photon spin
entanglement state.\\
\begin{eqnarray}
\chi_{23-1}&=&\frac{1}{\sqrt{15}}[2\chi_{-1}(s_{1z})\chi_{0}(s_{2z})\chi_{0}(s_{3z})+\chi_{-1}(s_{1z})\chi_{-1}(s_{2z})\chi_{1}(s_{3z})\nonumber\\
&&+\chi_{-1}(s_{1z})\chi_{1}(s_{2z})\chi_{-1}(s_{3z})+2\chi_{0}(s_{1z})\chi_{-1}(s_{2z})\chi_{0}(s_{3z})\nonumber\\&&+
2\chi_{0}(s_{1z})\chi_{0}(s_{2z})\chi_{-1}(s_{3z})+\chi_{1}(s_{1z})\chi_{-1}(s_{2z})\chi_{-1}(s_{3z})],
\end{eqnarray}
the spin state $\chi_{23-1}$ is the three-photon spin
entanglement state.\\
\begin{eqnarray}
\chi_{23-2}=\frac{1}{\sqrt{3}}[\chi_{-1}(s_{1z})\chi_{-1}(s_{2z})\chi_{0}(s_{3z})+\chi_{-1}(s_{1z})\chi_{0}(s_{2z})\chi_{-1}(s_{3z})+\chi_{0}(s_{1z})\chi_{-1}(s_{2z})\chi_{-1}(s_{3z})],
\end{eqnarray}
the spin state $\chi_{23-2}$ is the three-photon spin
entanglement state.\\
\begin{eqnarray}
\chi_{23-3}=\chi_{-1}(s_{1z})\chi_{-1}(s_{2z})\chi_{-1}(s_{3z}),
\end{eqnarray}
the spin state $\chi_{23-3}$ is not the three-photon spin
entanglement state.\\
\begin{eqnarray}
\chi_{222}=\frac{1}{\sqrt{6}}[\chi_{0}(s_{1z})\chi_{1}(s_{2z})\chi_{1}(s_{3z})+\chi_{1}(s_{1z})\chi_{0}(s_{2z})\chi_{1}(s_{3z})
-2\chi_{1}(s_{1z})\chi_{1}(s_{2z})\chi_{0}(s_{3z})],
\end{eqnarray}
the spin state $\chi_{222}$ is the three-photon spin
entanglement state.\\
\begin{eqnarray}
\chi_{221}&=&\frac{1}{\sqrt{12}}[\chi_{-1}(s_{1z})\chi_{1}(s_{2z})\chi_{1}(s_{3z})+2\chi_{0}(s_{1z})\chi_{0}(s_{2z})\chi_{1}(s_{3z})\nonumber\\
&&-\chi_{0}(s_{1z})\chi_{1}(s_{2z})\chi_{0}(s_{3z})+\chi_{1}(s_{1z})\chi_{-1}(s_{2z})\chi_{1}(s_{3z})\nonumber\\
&&-\chi_{1}(s_{1z})\chi_{0}(s_{2z})\chi_{0}(s_{3z})-2\chi_{1}(s_{1z})\chi_{1}(s_{2z})\chi_{-1}(s_{3z})],
\end{eqnarray}
the spin state $\chi_{221}$ is the three-photon spin
entanglement state.\\
\begin{eqnarray}
\chi_{220}&=&\frac{1}{2}[\chi_{-1}(s_{1z})\chi_{0}(s_{2z})\chi_{1}(s_{3z})+\chi_{0}(s_{1z})\chi_{-1}(s_{2z})\chi_{1}(s_{3z})\nonumber\\
&&-\chi_{0}(s_{1z})\chi_{1}(s_{2z})\chi_{-1}(s_{3z})-\chi_{1}(s_{1z})\chi_{0}(s_{2z})\chi_{-1}(s_{3z})],
\end{eqnarray}
the spin state $\chi_{220}$ is the three-photon spin
entanglement state.\\
\begin{eqnarray}
\chi_{22-1}&=&\frac{1}{\sqrt{12}}[2\chi_{-1}(s_{1z})\chi_{-1}(s_{2z})\chi_{1}(s_{3z})-\chi_{-1}(s_{1z})\chi_{1}(s_{2z})\chi_{-1}(s_{3z})\nonumber\\
&&-2\chi_{0}(s_{1z})\chi_{0}(s_{2z})\chi_{-1}(s_{3z})-\chi_{1}(s_{1z})\chi_{-1}(s_{2z})\chi_{-1}(s_{3z})\nonumber\\
&&+\chi_{-1}(s_{1z})\chi_{0}(s_{2z})\chi_{0}(s_{3z})+\chi_{0}(s_{1z})\chi_{-1}(s_{2z})\chi_{0}(s_{3z})],
\end{eqnarray}
the spin state $\chi_{22-1}$ is the three-photon spin
entanglement state.\\
\begin{eqnarray}
\chi_{22-2}&=&\frac{1}{\sqrt{6}}[2\chi_{-1}(s_{1z})\chi_{-1}(s_{2z})\chi_{0}(s_{3z})-\chi_{-1}(s_{1z})\chi_{0}(s_{2z})\chi_{-1}(s_{3z})\nonumber\\
&&-\chi_{0}(s_{1z})\chi_{-1}(s_{2z})\chi_{-1}(s_{3z})],
\end{eqnarray}
the spin state $\chi_{22-2}$ is the three-photon spin
entanglement state.\\
\begin{eqnarray}
\chi_{211}&=&\frac{1}{\sqrt{60}}[3\chi_{0}(s_{1z})\chi_{1}(s_{2z})\chi_{0}(s_{3z})+3\chi_{1}(s_{1z})\chi_{0}(s_{2z})\chi_{0}(s_{3z})\nonumber\\
&&-\chi_{1}(s_{1z})\chi_{-1}(s_{2z})\chi_{1}(s_{3z})-2\chi_{0}(s_{1z})\chi_{0}(s_{2z})\chi_{1}(s_{3z})\nonumber\\
&&-\chi_{-1}(s_{1z})\chi_{1}(s_{2z})\chi_{1}(s_{3z})-6\chi_{1}(s_{1z})\chi_{1}(s_{2z})\chi_{-1}(s_{3z})],
\end{eqnarray}
the spin state $\chi_{211}$ is the three-photon spin
entanglement state.\\
\begin{eqnarray}
\chi_{210}&=&\frac{1}{\sqrt{60}}[2\chi_{-1}(s_{1z})\chi_{1}(s_{2z})\chi_{0}(s_{3z})+4\chi_{0}(s_{1z})\chi_{0}(s_{2z})\chi_{0}(s_{3z})\nonumber\\
&&-3\chi_{0}(s_{1z})\chi_{1}(s_{2z})\chi_{-1}(s_{3z})+2\chi_{1}(s_{1z})\chi_{-1}(s_{2z})\chi_{0}(s_{3z})\nonumber\\
&&-3\chi_{1}(s_{1z})\chi_{0}(s_{2z})\chi_{-1}(s_{3z})],
\end{eqnarray}
the spin state $\chi_{220}$ is the three-photon spin
entanglement state.\\
\begin{eqnarray}
\chi_{21-1}&=&\frac{1}{\sqrt{60}}[3\chi_{0}(s_{1z})\chi_{-1}(s_{2z})\chi_{0}(s_{3z})+3\chi_{-1}(s_{1z})\chi_{0}(s_{2z})\chi_{0}(s_{3z})\nonumber\\
&&-6\chi_{-1}(s_{1z})\chi_{-1}(s_{2z})\chi_{1}(s_{3z})-\chi_{-1}(s_{1z})\chi_{1}(s_{2z})\chi_{-1}(s_{3z})\nonumber\\
&&-2\chi_{0}(s_{1z})\chi_{0}(s_{2z})\chi_{-1}(s_{3z})-\chi_{1}(s_{1z})\chi_{-1}(s_{2z})\chi_{-1}(s_{3z})],
\end{eqnarray}
the spin state $\chi_{22-1}$ is the three-photon spin
entanglement state.\\
\begin{eqnarray}
\chi_{121}&=&\frac{1}{\sqrt{4}}[\chi_{1}(s_{1z})\chi_{-1}(s_{2z})\chi_{1}(s_{3z})-\chi_{-1}(s_{1z})\chi_{1}(s_{2z})\chi_{1}(s_{3z})\nonumber\\
&&+\chi_{1}(s_{1z})\chi_{0}(s_{2z})\chi_{0}(s_{3z})-\chi_{0}(s_{1z})\chi_{1}(s_{2z})\chi_{0}(s_{3z})],
\end{eqnarray}
the spin state $\chi_{121}$ is the three-photon spin
entanglement state.\\
\begin{eqnarray}
\chi_{120}&=&\frac{1}{\sqrt{12}}[\chi_{0}(s_{1z})\chi_{-1}(s_{2z})\chi_{1}(s_{3z})-2\chi_{-1}(s_{1z})\chi_{1}(s_{2z})\chi_{0}(s_{3z})\nonumber\\
&&+2\chi_{1}(s_{1z})\chi_{-1}(s_{2z})\chi_{0}(s_{3z})-\chi_{-1}(s_{1z})\chi_{0}(s_{2z})\chi_{1}(s_{3z})\nonumber\\
&&+\chi_{1}(s_{1z})\chi_{0}(s_{2z})\chi_{-1}(s_{3z})-\chi_{0}(s_{1z})\chi_{1}(s_{2z})\chi_{-1}(s_{3z})],
\end{eqnarray}
the spin state $\chi_{120}$ is the three-photon spin
entanglement state.\\
\begin{eqnarray}
\chi_{12-1}&=&\frac{1}{2}[\chi_{0}(s_{1z})\chi_{-1}(s_{2z})\chi_{0}(s_{3z})-\chi_{-1}(s_{1z})\chi_{1}(s_{2z})\chi_{-1}(s_{3z})\nonumber\\
&&-\chi_{-1}(s_{1z})\chi_{0}(s_{2z})\chi_{0}(s_{3z})+\chi_{1}(s_{1z})\chi_{-1}(s_{2z})\chi_{-1}(s_{3z})],
\end{eqnarray}
the spin state $\chi_{12-1}$ is the three-photon spin
entanglement state.\\
\begin{eqnarray}
\chi_{12-2}
&=&\frac{1}{\sqrt{2}}[\chi_{0}(s_{1z})\chi_{-1}(s_{2z})\chi_{-1}(s_{3z})-\chi_{-1}(s_{1z})\chi_{0}(s_{2z})\chi_{-1}(s_{3z})],
\end{eqnarray}
the spin state $\chi_{12-2}$ is not the three-photon spin
entanglement state.\\
\begin{eqnarray}
\chi_{111}&=&\frac{1}{\sqrt{4}}[\chi_{1}(s_{1z})\chi_{-1}(s_{2z})\chi_{1}(s_{3z})-\chi_{-1}(s_{1z})\chi_{1}(s_{2z})\chi_{1}(s_{3z})
\nonumber\\&&-\chi_{1}(s_{1z})\chi_{0}(s_{2z})\chi_{0}(s_{3z})+\chi_{0}(s_{1z})\chi_{1}(s_{2z})\chi_{0}(s_{3z})],
\end{eqnarray}
the spin state $\chi_{111}$ is the three-photon spin
entanglement state.\\
\begin{eqnarray}
\chi_{110}&=&\frac{1}{\sqrt{4}}[\chi_{0}(s_{1z})\chi_{-1}(s_{2z})\chi_{1}(s_{3z})-\chi_{-1}(s_{1z})\chi_{0}(s_{2z})\chi_{1}(s_{3z})\nonumber\\
&&-\chi_{1}(s_{1z})\chi_{0}(s_{2z})\chi_{-1}(s_{3z})+\chi_{0}(s_{1z})\chi_{1}(s_{2z})\chi_{-1}(s_{3z})],
\end{eqnarray}
the spin state $\chi_{110}$ is the three-photon spin
entanglement state.\\
\begin{eqnarray}
\chi_{11-1}&=&\frac{1}{\sqrt{4}}[\chi_{-1}(s_{1z})\chi_{1}(s_{2z})\chi_{-1}(s_{3z})-\chi_{1}(s_{1z})\chi_{-1}(s_{2z})\chi_{-1}(s_{3z})\nonumber\\
&&+\chi_{0}(s_{1z})\chi_{1}(s_{2z})\chi_{0}(s_{3z})-\chi_{-1}(s_{1z})\chi_{0}(s_{2z})\chi_{0}(s_{3z})],
\end{eqnarray}
the spin state $\chi_{11-1}$ is the three-photon spin
entanglement state.\\
\begin{eqnarray}
\chi_{100}&=&\frac{1}{\sqrt{4}}[\chi_{0}(s_{1z})\chi_{-1}(s_{2z})\chi_{1}(s_{3z})-\chi_{-1}(s_{1z})\chi_{0}(s_{2z})\chi_{1}(s_{3z})\nonumber\\
&&+\chi_{1}(s_{1z})\chi_{0}(s_{2z})\chi_{-1}(s_{3z})-\chi_{0}(s_{1z})\chi_{1}(s_{2z})\chi_{-1}(s_{3z})],
\end{eqnarray}
the spin state $\chi_{100}$ is the three-photon spin
entanglement state.\\
\begin{eqnarray}
\chi_{011}&=&\frac{1}{\sqrt{3}}[\chi_1(s_{1z})\chi_{-1}(s_{2z})\chi_1(s_{3z})-\chi_0(s_{1z})\chi_{0}(s_{2z})\chi_1(s_{3z})
+\chi_{-1}(s_{1z})\chi_{1}(s_{2z})\chi_1(s_{3z})],
\end{eqnarray}
the spin state $\chi_{011}$ is not the three-photon spin
entanglement state.\\
\begin{eqnarray}
\chi_{010}=\frac{1}{\sqrt{3}}[\chi_1(s_{1z})\chi_{-1}(s_{2z})\chi_0(s_{3z})-\chi_0(s_{1z})\chi_{0}(s_{2z})\chi_0(s_{3z})
+\chi_{-1}(s_{1z})\chi_{1}(s_{2z})\chi_0(s_{3z})],
\end{eqnarray}
the spin state $\chi_{010}$ is not the three-photon spin
entanglement state.\\
\begin{eqnarray}
\chi_{01-1}=\frac{1}{\sqrt{3}}[\chi_1(s_{1z})\chi_{-1}(s_{2z})\chi_{-1}(s_{3z})-\chi_0(s_{1z})\chi_{0}(s_{2z})\chi_{-1}(s_{3z})
+\chi_{-1}(s_{1z})\chi_{1}(s_{2z})\chi_{-1}(s_{3z})].
\end{eqnarray}
The spin state $\chi_{01-1}$ is not the three-photon spin
entanglement state.\\

\vskip 5pt {\bf 6. The total entanglement between the electrons
and photons} \vskip 5pt

Through an example of two electrons and two photons, we shall give
their total entanglement states. For the Bose (Fermi) systems, the
total states include the space states and spin states and should
be symmetrical (antisymmetrical). So, the total entanglement
states of two electrons can be written as
\begin{eqnarray}
\psi^A(\vec{r}_1,\vec{r}_2,s_{1z},s_{2z},t)=\psi^S(\vec{r}_1,\vec{r}_2,t)\otimes
\chi_{00}^A(s_{1z},s_{2z}),
\end{eqnarray}
and
\begin{eqnarray}
\psi^A(\vec{r}_1,\vec{r}_2,s_{1z},s_{2z},t)=\psi^A(\vec{r}_1,\vec{r}_2,t)\otimes
\chi_{11}^S(s_{1z},s_{2z}),
\end{eqnarray}
where $\psi^S(\vec{r}_1,\vec{r}_2,t)$ and
$\psi^A(\vec{r}_1,\vec{r}_2,t)$ are space symmetrical and
antisymmetrical states, $\chi_{11}^S(s_{1z},s_{2z})$ and
$\chi_{00}^A(s_{1z},s_{2z})$ are spin symmetrical and
antisymmetrical entanglement states of two electrons, i.e., the
full entanglement states of two electrons are the direct product
of space states and spin entanglement states. When the space state
$\psi^S(\vec{r}_1,\vec{r}_2,t)\rightarrow 0$ or
$\psi^A(\vec{r}_1,\vec{r}_2,t)\rightarrow 0$, the entanglement of
two electrons should be disappeared even if they are in the spin
entanglement state $\chi_{00}^A(s_{1z},s_{2z})$ or
$\chi_{11}^S(s_{1z},s_{2z})$. So, the entanglement state of two
electrons only exist in the limited spatial scope. When they are
far away, the space state
$\psi^S(\vec{r}_1,\vec{r}_2,t)\rightarrow 0$ or
$\psi^A(\vec{r}_1,\vec{r}_2,t)\rightarrow 0$, the total
entanglement state $\psi^A(\vec{r}_1,\vec{r}_2,s_{1z},s_{2z},t)$
should approach to zero, the two electrons will not be in
entanglement state even if their spin state is entangled. In
experiments [24, 25], the authors have found the two electrons
entanglement exit within a limited space range, rather than an
arbitrary distance, which is agreement with the theory result.
Similarly, the total entanglement states of two photons can be
written as
\begin{eqnarray}
\psi^S(\vec{r}_1,\vec{r}_2,s_{1z},s_{2z},t)=\psi^S(\vec{r}_1,\vec{r}_2,t)\otimes
\chi_{2M_s}^S(s_{1z},s_{2z}), \hspace{0.3in}(M_s=1,0.-1)
\end{eqnarray}
\begin{eqnarray}
\psi^S(\vec{r}_1,\vec{r}_2,s_{1z},s_{2z},t)=\psi^A(\vec{r}_1,\vec{r}_2,t)\otimes
\chi_{1M_s}^A(s_{1z},s_{2z}), \hspace{0.3in}(M_s=1,0.-1)
\end{eqnarray}
and
\begin{eqnarray}
\psi^S(\vec{r}_1,\vec{r}_2,s_{1z},s_{2z},t)=\psi^S(\vec{r}_1,\vec{r}_2,t)\otimes
\chi_{00}^S(s_{1z},s_{2z}),
\end{eqnarray}
where $\psi^S(\vec{r}_1,\vec{r}_2,t)$ and
$\psi^A(\vec{r}_1,\vec{r}_2,t)$ are space symmetrical and
antisymmetrical states, $\chi_{2M_s}^S(s_{1z},s_{2z})$,
$\chi_{00}^S(s_{1z},s_{2z})$ and $\chi_{1M_s}^A(s_{1z},s_{2z})$
are spin symmetrical and antisymmetrical entanglement states of
two photons. Similarly, When two photons are far away, they will
not be in entanglement state even if their spin state is
entangled.

\newpage \vskip 5pt {\bf 7.
Conclusion} \vskip 5pt

In this paper, we have studied the entanglement mechanism of
electron-electron and photon-photon. With the spin-spin
interaction, we have given the spin entanglement states of two
electrons, three electrons, two photons, and found their total
entanglement states are relevant both space state and spin state.
When two electrons or two photons are far away, their entanglement
states should be disappeared even if their spin state is
entangled.

\end{document}